\documentclass[12pt]{article}
\usepackage{amsmath}
\usepackage{amsfonts}
\usepackage{graphicx,psfrag,epsf}
\usepackage{enumerate}
\usepackage{natbib}
\usepackage{chngcntr}
\usepackage{url} 
\usepackage{hyperref} 
\usepackage{authblk}
\usepackage{caption,xcolor}
\DeclareCaptionFont{blue}{\color{blue}}
\usepackage{booktabs}  

\newcommand{\blind}{0}

\usepackage[bottom=1.4in]{geometry}

\begin{document}

\def\spacingset#1{\renewcommand{\baselinestretch}%
{#1}\small\normalsize} \spacingset{1}


\if0\blind
{
  \title{\bf A Proposed Hybrid Effect Size Plus $p$-Value Criterion: A Comment on Goodman et al. (2019)}
  
  \author{Robin Tim Dreher \\
    Leona Hoffmann\\
    Arne Kramer-Sunderbrink \\
    Peter Pütz\footnote{Corresponding author: Peter Pütz, Bielefeld University, Faculty of Business Administration and Economics, Universitätsstr. 25, D-33615 Bielefeld, Germany. None of the authors has relevant or material financial interests that relate to the research described in this paper.} \\
    Robin Werner }
    \date{}

    \affil{\textit{Department of Economics, Bielefeld University}}
  \maketitle
} \fi

\bigskip
\begin{abstract}
In a recent simulation study, \citet{Goodman2019} compare several  methods  with  regard  to  their type I and type II error rates when considering a thick null hypothesis that includes all values that are practically equivalent to the point null hypothesis. They propose a hybrid
decision criterion only declaring a result ``significant'' if both a small $p$-value
and a sufficiently large effect size are obtained. We successfully verify
the results using our own software code in R and discuss an additional decision method that is tailored to maintain a pre-defined false positive rate. We confirm that the hybrid decision criterion has comparably low error rates in checkable settings but point out that the false discovery rate cannot be easily controlled by the researcher. Our analyses are readily accessible and customizable on \url{https://github.com/drehero/goodman-replication}.
(JEL: C10, C12)
\end{abstract}


\newpage
\spacingset{1.5}

\section{Introduction}
\label{sec:intro}

The discussion on the usefulness of null hypothesis significance testing and the correct use and interpretation of p-values has been hot in recent years. The \textit{American Statistical Association Statement on p-Values and Statistical Significance} \citep{wasserstein2016asa} was arguably the most prominent work on pitfalls in using $p$-values for statistical inference. It was followed by a symposium on statistical inference organized by the \textit{American Statistical Association} in 2017. The symposium led to a special issue of \textit{The American Statistician} titled "Moving to a World Beyond $p<0.05$" presenting many suggestions on how to separate signal from noise in data. In one of the included papers, \citet{Goodman2019}, henceforth GSK, compare several methods with regard to their type I and type II error rates when considering a thick null hypothesis, i.e., the null hypothesis includes all values that are practically equivalent to the point null hypothesis. They propose a hybrid decision criterion, the \textit{minimum effect size plus p-value} (MESP) that only leads to a rejection if a $p$-value lower than 5\% is obtained for the tested point null hypothesis and the observed effect size is large enough to be considered practically relevant. In a simulation study they compare different decision criteria and show good  overall performance of the MESP method. More specifically, MESP generally has rather low type I error rates, and has quite high type II error rates only in settings that the researcher can usually identify, namely low nominal power settings. 

In this comment, we successfully verify the GSK results using independently written software code in R. We extend the analyses by a Bayesian-motivated t-test tailored to maintain a pre-defined false positive rate. We also compute false discovery and false omission rates for different nominal power categories and show that the MESP approach has the undesirable property that the false discovery rate does not necessarily decrease for rising nominal power. Furthermore, we provide an openly available code repository that allows the scientific community to easily verify, amend and extend the analyses conducted in GSK and in this comment.

This comment proceeds as follows. Section \ref{sec:methods} explains the simulation setting and methods used by GSK and motivates a further decision criterion. Section \ref{sec:results} presents the results, while Section \ref{sec:conclusion} concludes.

\section{Simulation setting and decision methods}
\label{sec:methods}

In the recent discussion about the replication crisis, some authors suggested that the traditional focus on point hypotheses is part of the problem \citep[e.g.,][]{Greenland2019, McShane2019}: When testing the point null hypothesis $H_0^p: \mu = \mu_0$, even tiny, practically irrelevant effects can be ``statistically significant'' if the number of samples is big enough. Furthermore, problems in the experimental setup can easily lead to a small bias that is wrongly flagged as a statistically significant effect by the test. It was suggested to refrain from testing point hypotheses and instead testing whether an effect is big enough to be practically significant
\citep[e.g.,][]{betensky2019, Blume2019}.

In the following, the effect size deemed practically relevant by researchers in a certain scenario is referred to as $MPSD$ (Minimum Practically Significant Distance). Accordingly, the \emph{thick null hypothesis} is $H_0^t: |\mu-\mu_0| \le MPSD$ or, equivalently, $H_0^t:\mu \in I_0$, where $I_0=[\mu_0-MPSD,\mu_0+MPSD]$ is the \emph{thick null interval}.

\paragraph{Simulation}
We simulate 100,000 \emph{cases}.\footnote{In order to obtain robust results over different simulation runs also for subgroup analyses (e.g. Table \ref{tab:relativempsdcomparison}), we increase the number of simulated cases compared to GSK who simulate 10,000 cases.}
Each can be thought of as a fictional study testing whether some parameter is different from $\mu_0=100$. Formally, a case is characterized by a tuple $(\mu, \sigma, n, MPSD)$ describing the conditions of the study:
\begin{itemize}
    \item $\mu$, sampled uniformly from $\{75,\ldots,125\}$, is the population mean of our data.
    \item $\sigma$, sampled uniformly from $\{4,\ldots,60\}$, is the standard deviation of the population.
    \item $n$, sampled uniformly from $\{5,\ldots,100\}$, is the sample size.
    \item $MPSD$, sampled uniformly from $\{2,\ldots,20\}$, is the minimal effect size deemed practically relevant by researchers in our fictional scenario.
\end{itemize}

For every case, we save if $|\mu - \mu_0| > MPSD$ (i.e., if $H_0^t$ is false). We sample the data $x_1,\ldots,x_n \overset{i.i.d.}{\sim} \mathcal{N}(\mu, \sigma)$ and save the results of the decision methods described below based on that sample.

\subsection{Original decision methods}

GSK compare five different decision methods:

\paragraph{Conventional} The first method is a $t$-test for the point null hypothesis that $\mu = \mu_0$ with $\alpha = 0.05$. Note that this $\alpha$ is usually the desired false positive rate (the type I error): If the point null hypothesis is true it should not be falsely rejected by the test more than 5\% of the time. However, since the test is used here to decide whether the thick $H_0^t$, not $H_0^p$, should be rejected, the false positive rate will exceed 5\% for $MPSD>0$. 

\paragraph{Small-alpha} The second method is a $t$-test as described above with $\alpha = 0.005$ instead of $0.05$.
Lowering $\alpha$ from 0.05 to 0.005 was prominently recommended to lower the rate of false positive results and irreproducible research findings \citep{benjamin2017}. However, as discussed above, $\alpha$ is used to control the false positive rate with respect to $H_0^p$, not $H_0^t$.

\paragraph{Distance-only} The third method rejects the null hypothesis if the observed effect size is at least as big as the minimum practically significant distance, i.e., $|\bar{x}-\mu_0| \ge MPSD$.

\paragraph{MESP} GSK propose a new decision method called ``Decision by Minimum Effect Size Plus $p$-value'' that is a conjunction of the conventional and the distance-only method, i.e., the null hypothesis is rejected only if both methods would reject it, if both the $p$-value of the $t$-test is smaller than 0.05 and the observed effect size is practically significant.

\paragraph{Interval-based} The final method considered by GSK rejects the null only if the thick null interval $I_0$ and the 95\% confidence interval centered around the observed sample mean $\bar{x}$ do not intersect.\footnote{The interval-based method is equivalent to a two-sided decision criterion presented in \cite{betensky2019} for rejecting a thick null hypothesis. However, \cite{betensky2019} allows the interval-based method also to be used to confirm thick null hypotheses. In this regard, the method applied by \cite{betensky2019} can be considered as a composite approach that combines the interval-based method presented in GSK and equivalence testing using two one-sided tests \citep{schuirmann}. The latter is designed to confirm the hypothesis that an effect is not larger than some practically meaningful equivalence bounds, i.e. the boundaries of the thick null. The Bayesian counterpart to \cite{betensky2019} is the HDI-ROPE procedure \citep{kruschke2011} that also allows for rejection and acceptance of a thick null hypothesis. Both \cite{betensky2019} and \cite{kruschke2011} deem results inconclusive if the estimated interval (confidence or credible interval, respectively) is neither fully inside nor fully outside of the thick null hypothesis.}

\subsection{Thick \textit{t}-test}
\label{sec:methods_t}

In this section we present a sixth decision method that we call the \emph{thick $t$-test}. The idea is to base the decision to reject the thick null hypothesis $H_0^t$ on the probability of observing an effect that is more extreme with respect to $\mu_0$ than the one actually observed if $H_0^t$ was true: 
\begin{align}\label{eq:thick}
\begin{split}
    p &= P(|\bar{X}-\mu_0| > |\bar{x}-\mu_0|\ |\ H_0^t)\\
      &= \sum_{\dot{\mu} = \mu_0- MPSD}^{\mu_0+ MPSD} P(|\bar{X}-\mu_0| > |\bar{x}-\mu_0|\ |\ \mu=\dot{\mu}) P(\mu = \dot{\mu}\ |\ H_0^t).
      \end{split}
\end{align}
Note that $\mu$ is an integer in our simulation. If $\mu$ is continuous, an integral has to be used instead of the sum. Our code contains details on how to compute $p$ in either case.

The probability in Equation (\ref{eq:thick}) is known as the \emph{prior predictive $p$-value} popularized by \citet{box1980} \citep[see also][for a discussion on this and related concepts for composite null hypotheses]{berger2000}.
For $MPSD=0$, i.e., $H_0^t \equiv H_0^p$, $p$ is simply the $p$-value of a two-sided point hypothesis test. In that sense, $p$ as defined above is a generalization of familiar $p$-values to thick null hypotheses and the same guarantee regarding the false positive rate from point hypothesis tests holds for the thick $t$-test: If we decide to reject $H_0^t$ whenever $p < \alpha = 0.05$, the probability to report an effect even though the real effect is practically irrelevant is 5\%. This is because, just like the familiar $p$-value in a two-sided point hypothesis test, $p$ is uniformly distributed on $[0,1]$ under $H_0^t$, i.e., $P(p \le \alpha\ |\ H_0^t) = \alpha$. To be precise, due to the discrete distribution of $\mu$ in the simulation, $p$ is only asymptotically uniformly distributed, and hence $P(p \le \alpha\ |\ H_0^t) = \alpha$ only for $n \to \infty$.

However, we need to make an assumption about the distribution of $\mu$ under $H_0^t$ and only under this assumption, the guarantee holds. Since we are doing a simulation we know that $\mu$ is uniformly distributed, i.e., $P(\mu = \dot{\mu}\ |\ H_0^t) = \frac{1}{2MPSD+1}$ for $\dot{\mu}\in I_0$. In a real-world experiment on the other hand, the distribution under $H_0^t$ used in the thick $t$-test is an expression of the researcher's prior knowledge or assumptions about $\mu$. For example, if a researcher has no or decides to ignore prior knowledge, they should opt for a flat prior for $\mu$ on the thick null interval. If a researcher believes that small effects are more likely than bigger ones, they could, for example, choose a truncated normal distribution.  Aside from or additional to a researcher's beliefs, prior knowledge on a parameter of interest may be gathered from expert elicitation \citep[e.g.,][]{albert2012} or previous studies in a specific research field \citep[e.g.,][see Appendix \ref{app:prior_meta} for an example]{schulz2021}. 

We believe that this method is useful for a number of reasons: Firstly, only two of the five original methods tested in GSK (MESP and the interval-based method) take both the $MPSD$ and the standard error of the estimated mean $\hat{\mu} = \bar{x}$ into account. Therefore, it benefits the comparison to add a decision method that uses all available information.

Secondly, it presents a framework to better understand most of the other methods: The conventional and the small-alpha method can be thought of as thick $t$-tests that assume $P(\mu = \mu_0 |\ H_0^t) = 1$.
For a given observed effect, this distributional choice produces the smallest $p$-values among all possible thick $t$-tests with a symmetric prior. Thus, $H^t_0$ is more often rejected by the test and may even be rejected if the observed effect is inside the thick null interval. Hence, we expect the false positive rate of these methods to be higher than $\alpha$ (5\% and 0.5\%, respectively).
The interval-based method on the other hand is approximately equivalent (except for a term that is usually negligible) to the thick $t$-test that assumes $P(\mu = \mu_c \ |\ H_0^t) = 1$, where $\mu_c$ is the value in the thick null interval that is closest to the observed mean $\bar{x}$. 
For a given observed effect, this distributional choice produces the largest $p$-values and is therefore the most conservative decision with respect to rejecting $H_0^t$ among all possible thick $t$-tests. Hence, we expect the false positive rate to be below 5\% and the false negative rate to be high for this method. Note that the thick $t$-test with $P(\mu = \mu_c \ |\ H_0^t) = 1$ is equivalent to choosing $p = \sup_{\dot{\mu}\in I_0} P(|\bar{X}-\mu_0| > |\bar{x}-\mu_0|\ |\ \mu=\dot{\mu})$. Taking the supremum instead of integrating $\dot{\mu}$ out is the traditional way of dealing with composite null hypotheses \citep[see for example][p. 217]{bickel2015}, e.g., when conducting a one-sided $t$-test.
Finally, the distance-only method can be thought of as a thick $t$-test that assumes the same distribution as the interval-based method but with $\alpha=50\%$. Hence, we expect the false positive rate to be higher than that of the interval-based method but well below 50\% since the assumed distribution is overly conservative.

Thirdly, the thick $t$-test could be used as a decision method in its own right. The $\alpha$ parameter allows controlling the risk of incorrectly rejecting a true null hypothesis directly and more accurately than the overly conservative interval-based method. Of course, this requires that the assumption about the distribution of the population mean when there is no practically relevant effect is at least approximately true. We don't believe that the need to make this assumption is necessarily a disadvantage compared with the other methods since one would have to make similar assumptions about the distribution of $\mu$ under $H_0^t$ if one wanted to derive any guarantees regarding their false positive rates with respect to the thick null hypothesis.

In the Appendix, we investigate the sensitivity of the thick $t$-test with regard to a mismatch between prior and true simulation distribution (Appendix \ref{app:prior_simu}) and compare its ability to control the false positive rate by choosing $\alpha$ to the other decision methods (Appendix \ref{app:alpha}). 

\section{Results}
\label{sec:results}

\begin{figure}[t!]
    \centering
    \includegraphics[width=0.7\textwidth]{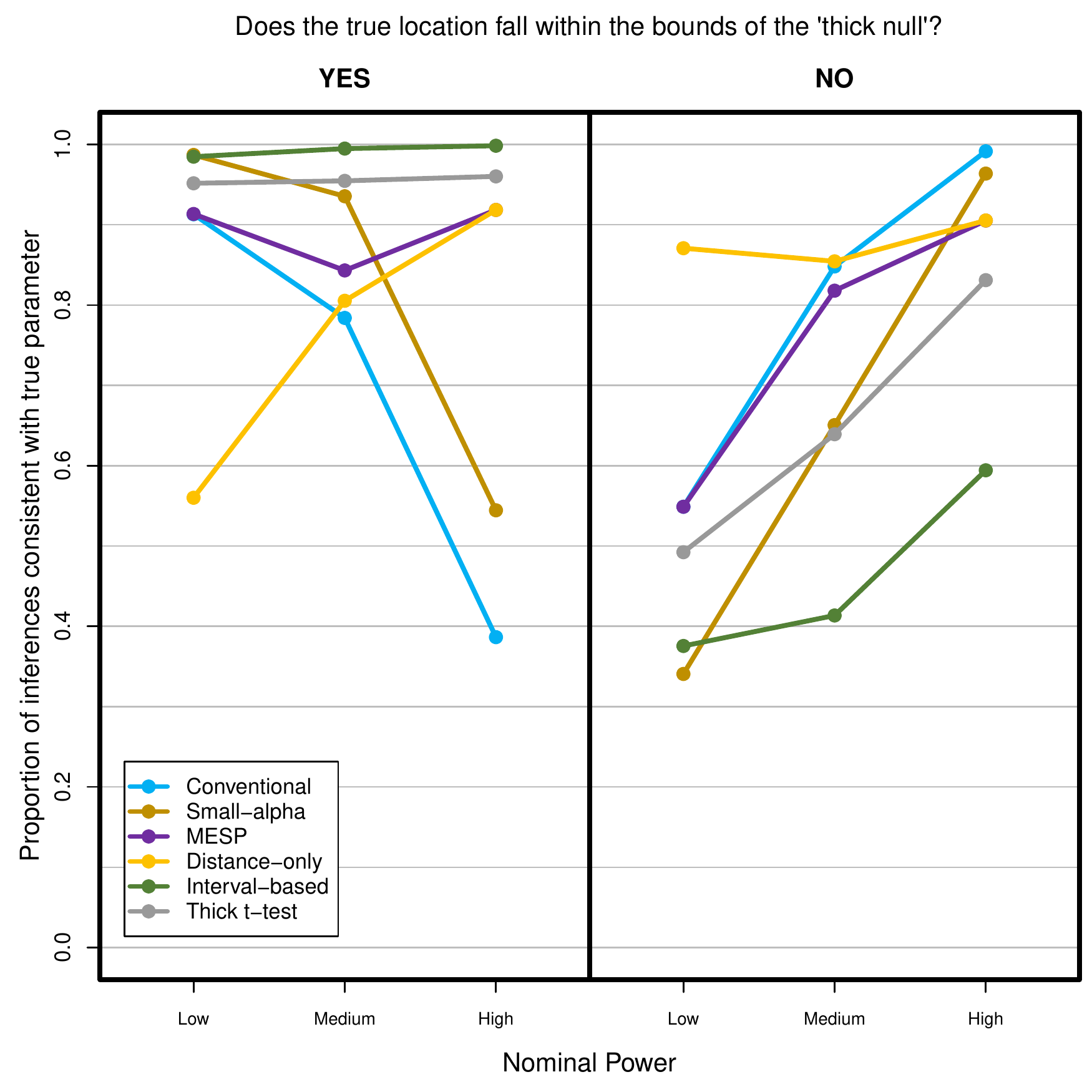}
    \caption{Graph of impact of nominal power, method, and true location of the null on inference success.}
    \label{fig:power_comparison}
\end{figure}

Table \ref{tab:power_comparison} was created based on Table 2 of GSK. The different methods are compared in terms of their overall inference success rate, controlled for the nominal power. The nominal power for a case in the simulation relies on the power calculation for a one-sample z-test for the mean using $\alpha=0.05$, the case’s true population standard deviation $\sigma$, sample size $n$ and minimum detectable difference $MPSD$. Table \ref{tab:power_comparison} also distinguishes between the true mean falling within and beyond the thick null. In addition to the methods from GSK, the thick $t$-test introduced in the last section is considered. The table is visualized in Figure \ref{fig:power_comparison}.

\begin{table}[t!]
\caption{Impact of nominal power, method, and true location of the population mean on inference success.}
\label{tab:power_comparison}
\resizebox{\columnwidth}{!}{
\begin{tabular}{lcccccccc} 
\toprule
& & & \multicolumn{6}{c}{\shortstack[]{Inference success rates of each method for each combination of true location and nominal power}} \\
\cmidrule(lr){4-9}
\shortstack[l]{Does the true location \\ fall within the bounds \\ of the thick null?} & Nominal power & \shortstack[]{Number of\\ simulated cases} & Conventional & Small-alpha & MESP & Distance-only & Interval-based & Thick $t$-test\\
 \midrule
  Yes (45,193 cases) & $\geq 0.80$ & 23,869 & 37.6\% & 53.3\% & 90.7\% & 90.7\% & 99.7\% & 95.2\% \\ 
  &  $0.30$ to $0.80$ & 12,920 & 76.3\% & 92.9\% & 82.6\% & 78.8\% & 99.5\% & 95.1\% \\ 
  & $< 0.30$ & 8,404 & 90.7\% & 98.4\% & 90.7\% & 55.7\% & 98.5\% & 94.7\% \\ 
  No (54,807 cases) &  $\geq 0.80$ & 17,674 & 99.4\% & 96.7\% & 92.4\% & 92.4\% & 62.6\% & 85.8\% \\ 
  & $0.30$ to $0.80$ & 14,507 & 85.5\% & 66.2\% & 83.0\% & 86.8\% & 42.9\% & 65.3\% \\ 
  & $< 0.30$ & 22,626 & 56.3\% & 35.3\% & 56.3\% & 88.0\% & 38.7\% & 50.6\% \\    
 \bottomrule
\end{tabular}
}
\end{table}

The top half of Table \ref{tab:power_comparison} and the left half of Figure 1 show the success rates of the methods when the true parameter lies within the thick null interval. The interval-based method performs best in this case, as already seen in GSK. The new method of the thick $t$-test falsely rejects $H_0^t$ approximately 5\% of the time across all nominal power categories. The success rates of the other methods except for the interval-based method are lower in at least one case and depend on the nominal power. The MESP method exhibits overall moderate error rates.

The bottom half of Table \ref{tab:power_comparison} and the right half of Figure \ref{fig:power_comparison} show the success rates of the methods when the real parameter lies outside the null interval. The previously best method performs worst in this case. Overall, the distance-only method has high inference success rates and is topped only at high nominal power by the conventional method and the small-alpha method. However, these two methods have a much worse success rate for low nominal power. The newly considered thick $t$-test performs moderately. It is neither one of the best methods at high nominal power, nor one of the worst methods at low nominal power. MESP performs reasonably well for higher nominal power and exhibits low inference success only in the low nominal power setting.

Overall, the described Table \ref{tab:power_comparison} and Figure \ref{fig:power_comparison} very much resemble the corresponding illustrations in GSK. The thick $t$-test method performs very well if the true location falls within the bounds of the thick null. MESP has generally acceptably low error rates except for low nominal power settings when the thick null is not true. As a composite measure of the conventional and the distance-only method it has inference success rates similar to the former when the nominal power is low and similar to the latter when the nominal power is high. The reason for this behavior is intuitive: The $p$-value computed for $H_0^p$ only plays an important role for the MESP decision if effect sizes need to be really large to lead to a rejection of $H_0^t$. Conversely, in high nominal power settings $p$-values are generally low and MESP's decision is only dependent on the distance criterion. An interesting property of MESP is that due to its construction as a composite measure it has lowest inference success for medium nominal power if the thick null holds.

\begin{table}[t!]
\caption{Impact of relative $MPSD$ and method on inference success.}
\label{tab:relativempsdcomparison}
\resizebox{\columnwidth}{!}{
\begin{tabular}{lccccccccc}
\toprule
& & & \multicolumn{6}{c}{\shortstack[]{Inference success rates of each method for each combination of true location and relative $MPSD$}} \\
\cmidrule(lr){4-9}
\shortstack[l]{Does the true location \\ fall within the bounds \\ of the thick null?} & 
\shortstack[]{Decile\textsuperscript{a} for \\ $MPSD / \sigma$ } & 
\shortstack[]{Number of \\ simulated cases} &
Conventional & Small-alpha & MESP & Distance-only & Interval-based & Thick $t$-test\\ 
 \midrule
  Yes (45,193 cases) & 1 & 1,355 & 93.6\% & 98.9\% & 93.6\% & 38.5\% & 97.5\% & 94.8\% \\ 
  & 2 & 2,459 & 90.8\% & 98.3\% & 90.8\% & 54.7\% & 98.9\% & 94.6\% \\ 
  & 3 & 3,305 & 85.4\% & 96.6\% & 85.8\% & 66.4\% & 98.9\% & 94.5\% \\ 
  & 4 & 4,297 & 80.1\% & 94.8\% & 83.6\% & 74.1\% & 99.3\% & 95.3\% \\ 
  & 5 & 5,147 & 75.1\% & 90.9\% & 85.2\% & 79.0\% & 99.4\% & 94.5\% \\ 
  & 6 & 5,633 & 68.0\% & 85.8\% & 84.9\% & 80.8\% & 99.5\% & 95.3\% \\ 
  & 7 & 5,605 & 58.3\% & 78.1\% & 86.4\% & 84.0\% & 99.5\% & 95.1\% \\ 
  & 8 & 5,617 & 48.6\% & 65.8\% & 89.3\% & 88.1\% & 99.7\% & 95.3\% \\ 
  & 9 & 5,730 & 34.6\% & 49.2\% & 91.6\% & 91.4\% & 99.6\% & 95.2\% \\ 
  & 10 & 6,045 & 16.9\% & 25.7\% & 95.1\% & 95.1\% & 99.9\% & 95.3\% \\ 
  No (54,807 cases) & 1 & 8,645 & 54.6\% & 34.7\% & 54.6\% & 91.8\% & 43.9\% & 51.9\% \\ 
  & 2 & 7,541 & 63.3\% & 43.6\% & 63.3\% & 89.7\% & 43.5\% & 57.0\% \\ 
  & 3 & 6,695 & 71.1\% & 50.4\% & 71.0\% & 87.6\% & 41.8\% & 58.8\% \\ 
  & 4 & 5,703 & 75.9\% & 57.1\% & 74.6\% & 85.4\% & 39.2\% & 60.0\% \\ 
  & 5 & 4,853 & 81.3\% & 64.0\% & 77.2\% & 85.1\% & 38.2\% & 61.3\% \\ 
  & 6 & 4,367 & 86.4\% & 71.9\% & 80.2\% & 84.7\% & 40.0\% & 64.1\% \\ 
  & 7 & 4,395 & 91.6\% & 80.8\% & 83.8\% & 86.9\% & 46.1\% & 70.7\% \\ 
  & 8 & 4,383 & 96.0\% & 90.6\% & 88.8\% & 90\% & 55.2\% & 80.1\% \\ 
  & 9 & 4,270 & 98.4\% & 95.5\% & 92.5\% & 93.1\% & 64.6\% & 87.7\% \\ 
  & 10 & 3,955 & 99.9\% & 99.2\% & 97.1\% & 97.1\% & 79.5\% & 95.9\% \\ 
 \bottomrule
\multicolumn{9}{l}{\textsuperscript{a}\footnotesize{These ranges of values for MPSD/$\sigma$ correspond to the deciles:}} \\
\footnotesize{
\begingroup
\setlength{\tabcolsep}{10pt} 
\renewcommand{\arraystretch}{0.6} 
    \begin{tabular}{l r}
        1: 0.033–0.103 & 6: 0.345–0.419 \\
        2: 0.103–0.167 & 7: 0.419–0.533 \\
        3: 0.167–0.224 & 8: 0.533–0.739 \\
        4: 0.224–0.286 & 9: 0.739–1.214 \\
        5: 0.286–0.345 & 10: 1.214–5.000
    \end{tabular}
\endgroup
}
\end{tabular}
}
\end{table}

\begin{figure}[t!]
    \centering
    \includegraphics[width=0.7\textwidth]{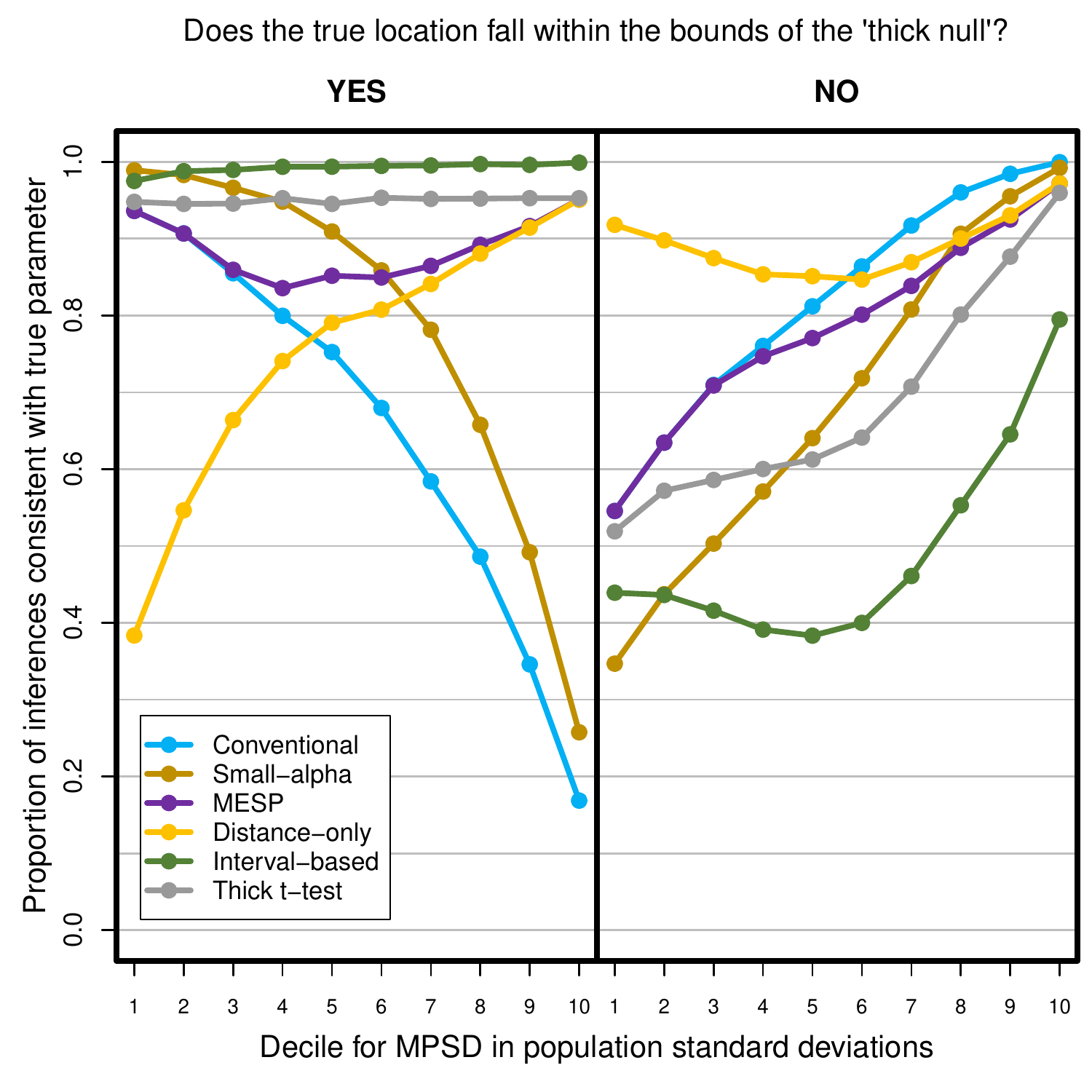}
    \caption{Impact of relative $MPSD$ and method on inference success.}
    \label{fig:mpsd_comparison}
\end{figure}

Following Table 3 of GSK, Table \ref{tab:relativempsdcomparison} shows inference success rates depending on deciles of the relative $MPSD$ which is defined as $MPSD/\sigma$. As $MPSD$ and $\sigma$ are drawn from respective ranges of integers for the simulation of cases, ties in the relative $MPSD$ occur. When dividing the simulated cases into deciles, we add a tiny value $\epsilon \sim \mathcal{N}(0, 10^{-10})$ to the relative $MPSD$ to distribute ties on the boundary of two deciles evenly to those deciles. 
The table is divided into two parts. The upper one shows success rates when the true location falls within the bounds of the thick null while the lower one shows success rates when the true location does not fall within the bounds of the thick null. Figure \ref{fig:mpsd_comparison} visualizes Table \ref{tab:relativempsdcomparison}.

Looking at the upper half of the table one can observe that the interval-based and the thick $t$-test method both have low error rates with the interval-based method showing almost no inference failure for high relative $MPSD$s. The MESP method has the best success rates for low and high deciles and performs moderately for medium deciles. For the distance-only method the success rate increases considerably the higher the relative $MPSD$ becomes. In contrast to that stand the conventional and the small-alpha method. They perform best for the low relative $MPSD$ values and fall off the higher it becomes. 

If the thick null is false, the conventional and small-alpha method, MESP as well as the thick $t$-test perform better the higher the relative $MPSD$ becomes. The interval-based method has considerably worse success rates than the other methods in all but the lowest deciles. The distance-only method has decent success rates, especially for low and high deciles. The MESP method performs nearly as good as the conventional $t$-test with slightly worse performance in medium and higher deciles. It exhibits inference success rates below 70\% only in settings where the relative $MPSD$ is low. In practice, such a setting could be recognized by the researcher before conducting a study if a reliable guess on the population's standard deviation is available.

Overall, these numbers confirm the results in GSK regarding the success rates for different ranges of the relative $MPSD$. As already shown in Table \ref{tab:power_comparison} and Figure \ref{fig:power_comparison}, the added thick $t$-test falsely rejects the thick null less often compared to other methods tailored to thick null hypotheses such as the MESP and the distance-only method, but has lower inference success rates if the thick null does not hold.

\begin{table}[t!]
\caption{Error rates and inference success rate for each method.}
\label{tab:error_rates}
\resizebox{\columnwidth}{!}{
\begin{tabular}{lcccccc}
\toprule
& Conventional & Small-alpha & MESP & Distance-only & Interval-based & Thick $t$-test \\ 
  \cmidrule(lr){2-7}
  False positive rate & 41.4\% & 27.0\% & 11.6\% & 19.2\% & 0.6\% & 4.9\% \\ 
  False negative rate & 22.1\% & 36.7\% & 25.0\% & 10.9\% & 52.5\% & 34.2\% \\ 
  False discovery rate & 30.5\% & 26.0\% & 11.3\% & 15.1\% & 1.0\% & 5.8\% \\ 
  False omission rate & 31.4\% & 37.9\% & 25.5\% & 14.1\% & 39.0\% & 30.4\% \\
  \addlinespace
  Inference success rate & 69.2\% & 67.7\% & 81.1\% & 85.3\% & 71.0\% & 79.0\% \\
  \bottomrule
\end{tabular}
}
\end{table}

Table \ref{tab:error_rates} and Table \ref{tab:fdr-for} are inspired by Figure A7 of GSK. Table \ref{tab:error_rates} shows error rates as well as overall inference success rates, i.e., the share of correct inference decisions over all simulations, with respect to the thick null.
With an inference success rate of 69.2\% for the conventional and 67.7\% for the small-alpha variant, the point $t$-tests together with the interval-based method (71.0\%) have the lowest inference success rate. With 85.3\% the distance-only method has the highest overall inference success rate. Of course, in another simulation setting with the thick null being true more often than 45.2\% as in our setting (Table \ref{tab:power_comparison}), the picture might look different.
Table \ref{tab:error_rates} also shows that the conventional $t$-test has the highest false positive rate (41.4\%) with respect to the thick null.
Lowering the alpha of the $t$-test to 0.005 reduces the false positive rate to 27.0\% at the expense of a higher false negative rate.

Combining the inferences of the conventional $t$-test and the distance-only method results in an inference success rate of 81.1\% for the MESP method with a false positive rate of 11.6\% and a false negative rate of 25.0\%.
The interval-based method has the lowest false positive rate (0.6\%) but with 52.5\% also the highest false negative rate of all six methods.
Due to its construction the thick $t$-test yields a false positive rate of 4.9\% and has a false negative rate that is 18.3 percentage points lower compared to the interval-based method. 

The false discovery rate is important from a practical perspective, as it is defined as the share of false positive findings among all positive findings (rejections of the null hypothesis). In the context of a thick null hypothesis, this is the probability that there is actually no practically relevant effect in the population if the decision criterion indicates that there is one. In our simulation, the false discovery rate is lowest for the interval-based method, while the false omission rate as its counterpart is lowest for the distance-only method. While the false discovery and false omission rate computed here are dependent on the share of true thick null hypotheses and can be derived from Table \ref{tab:power_comparison}, one would have to decide on a prior probability of the thick null being true for a single study in practice. 

Table \ref{tab:fdr-for} compares the six methods regarding the false discovery rate and the false omission rate in more detail. In this table, a distinction is made between high, medium and low nominal power. As the rates are crucially dependent on the share of thick null hypotheses being true, the rates are normalized by using rates to compute them instead of absolute numbers, i.e., \begin{align*}
    \text{normalized false discovery rate} &= \frac{\text{false positive rate}}{\text{false positive rate} + \text{true positive rate}} \text{  and} \\
     \text{normalized false omission rate} &= \frac{\text{false negative rate}}{\text{false negative rate} + \text{true negative rate}},
\end{align*} where the true/false positive/negative rates can be derived from Table \ref{tab:power_comparison}. 
To this end, comparisons between different nominal power categories are allowed and it can be investigated whether methods improve with increasing nominal power. All rates should only be interpreted relative to each other since, in general, the false discovery rate decreases and the false omission rate increases with more true locations falling beyond the thick null. Thus, a different simulation setting, e.g., with more true locations falling beyond the thick null, would lead to different results.  

The interval-based  method has false discovery rates which are well below 5\% for all nominal power categories. The thick $t$-test method also exhibits overall low false discovery rates. The first three methods shown in Table \ref{tab:fdr-for} do not have the generally desirable property of a decreasing false discovery rate for rising nominal power. While the MESP has overall medium false discovery rates, the false discovery rate rises when moving from low to medium nominal power. 
The reason is the aforementioned construction of the MESP as a composite measure of the conventional and the distance-only method which have contradicting false discovery rates when the nominal power increases.

Regarding the false omission rate, a consistent structure can be recognized across all methods: The false omission rate increases with decreasing nominal power. For high nominal power, the value is lowest for the conventional method (1.6\%) and highest for the interval-based method (27.3\%). For low nominal power, the distance-only method shows the lowest value of 17.7\%.

When considering the false discovery rate and the false omission rate, there is no clear best method even when disregarding the low nominal power category. For medium and high nominal power, many methods have false discovery rates that arguably exceed desirable values. The interval-based method works best in this regard at the expense of a considerable false omission rate even in the highest nominal power category. To this end, the individual context of a study plays a crucial role in the choice of the appropriate decision criterion.

\begin{table}[t!]
\caption{(Normalized) false discovery rate and (normalized) false omission rate depending on nominal power.}
\label{tab:fdr-for}
\resizebox{\columnwidth}{!}{
\begin{tabular}{lccccccccc}
  \toprule
  & Nominal Power & \shortstack[c]{Number of\\ sim. cases} & Conventional & \shortstack[c]{Small-\\ alpha} & MESP & \shortstack[c]{Distance- \\only} & \shortstack[c]{Interval-\\ based} & \shortstack[c]{Thick\\ $t$-test} \\ 
  \cmidrule(lr){2-9}
  (Normalized) false discovery rate & $\geq$ 0.80 & 41,543 & 38.6\% & 32.6\% & 9.1\% & 9.1\% & 0.5\% & 5.3\% \\ 
  & 0.30 to 0.80 & 27,427 & 21.7\% & 9.7\% & 17.3\% & 19.7\% & 1.2\% & 7\% \\ 
  & $<$ 0.30 & 31,030 & 14.2\% & 4.2\% & 14.2\% & 33.5\% & 3.7\% & 9.5\% \\ 
  (Normalized) false omission rate & $\geq$ 0.80 & 41,543 & 1.6\% & 5.8\% & 7.7\% & 7.7\% & 27.3\% & 13\% \\ 
  & 0.30 to 0.80 & 27,427 & 16\% & 26.7\% & 17\% & 14.4\% & 36.5\% & 26.7\% \\ 
  & $<$ 0.30 & 31,030 & 32.5\% & 39.6\% & 32.5\% & 17.7\% & 38.3\% & 34.3\% \\  
  \bottomrule
\end{tabular}
}
\end{table}

\section{Conclusion and Discussion}
\label{sec:conclusion}
In this comment, we confirmed the results of GSK using our own software code written in R that enables the easy implementation of further decision criteria. We added such a decision criterion that allows to control for a pre-defined type I error rate with respect to a thick null hypothesis.

We confirmed that the MESP method as proposed by GSK has comparably low type I and type II error rates with respect to a thick null hypothesis in settings the researcher can check for. Only in low nominal power settings the MESP method has difficulties to detect a practically relevant effect and, accordingly, has a quite high false omission rate. The false discovery rate is moderate in comparison with other decision criteria but has the undesirable property that it does not decrease monotonically with rising nominal power (assuming a fixed prior probability for the thick null being true). If the researcher's aim is to keep the false discovery rate low, the MESP should only be used in high nominal power settings.  

The MESP method could also be applied for \emph{point} null hypothesis testing: As it augments the conventional $t$-test by taking the minimum practically significant effect size into account, MESP is a simple and effective approach for adding an additional layer of protection against false positives without strongly increasing the false negative rate. However, if researchers are serious about testing whether the \emph{thick} null hypothesis is true, MESP lacks a parameter to adapt the method to contexts with different costs of false negatives and positives with respect to the thick null hypothesis. In contrast, the interval-based method and the thick $t$-test (assuming the chosen distribution of $\mu\ |\ H_0^t$ is approximately true) provide such a parameter with $\alpha$ being the upper bound of the false positive rate or approximately equal to it, respectively. 

The performance of the decision criteria generally depends on whether the thick null hypothesis is true or not. In practice, one can make an informed guess on this probability, i.e., specifying a prior probability, or select the decision criterion according to the estimated nominal power and the type of error one wants to avoid. Likewise, for hypothesis testing it is recommendable to specify and publish the (thick) null hypothesis, planned analyses methods, and the decision criterion before conducting the study to reduce options for $p$-hacking and increase transparency in science. Irrespective of the decision criterion used, every empirical study is subject to limitations and assumptions and thus uncertainty beyond sampling noise. While an extensive debate on statistical inference including dos and don'ts as well as merits and pitfalls can be found in \cite{wasserstein2019} and the corresponding issue of the \textit{The American Statistician}, we conclude by citing a generic piece of advice to the research community using statistics: “Accept uncertainty. Be thoughtful, open, and modest” \citep{wasserstein2019}.

\bibliographystyle{Chicago}
\bibliography{bibliography}

\newpage
\setcounter{page}{1}
\pagenumbering{Alph}
\appendix
\counterwithin{figure}{section}
\counterwithin{table}{section}
\section{Appendix\label{supp}}
\subsection{Practical example of choosing a prior based on previous research  \label{app:prior_meta}}
Choosing the prior for the thick $t$-test based on previous research can be exemplified by using data from \cite{van2019} that comprise 83 meta-analyses published in \textit{Psychological Bulletin} including 366 homogeneous subsets of at least five effect sizes from primary studies. All primary effect sizes and their sampling variances are transformed to Cohen's $d$ and random-effects meta-analyses are used to estimate overall meta-analytic effect sizes, i.e., 366 true effect sizes are estimated. Following the convention that absolute values of Cohen's $d$ below 0.2 constitute small (less relevant) effects, one could consider the distribution of the 164 resulting meta-analytic effect size estimates with $-0.2 < d < 0.2$ as the prior distribution of $\mu$ under the thick null hypothesis for psychological studies. This approximation seems justified, as meta-analytic estimates are usually based on large samples and thus expected to estimate the underlying true effects with high precision (although issues such as publication bias should be considered).
\begin{figure}[t!]
    \centering
    \includegraphics[width=0.65\textwidth]{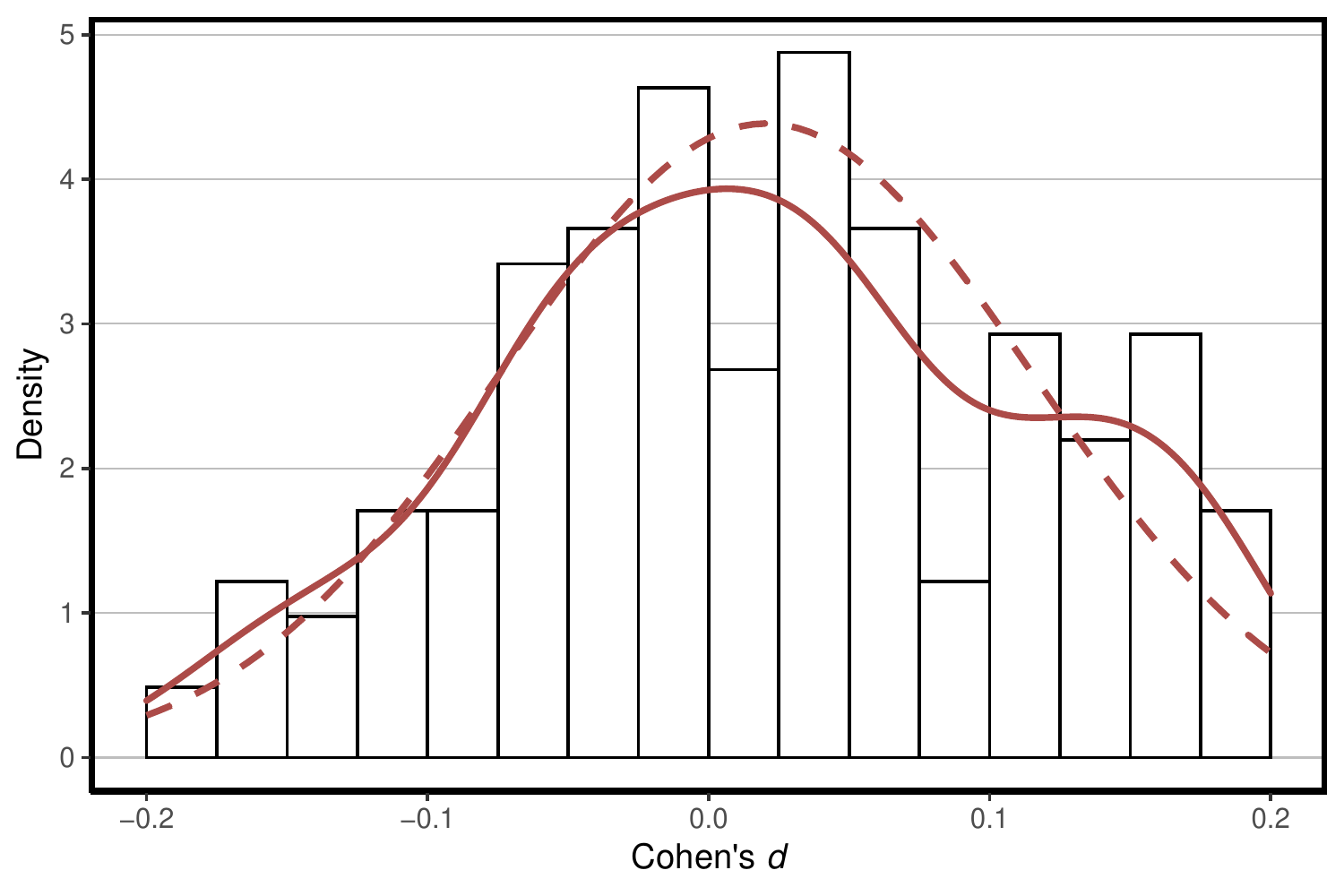}
    \caption{Estimated distribution of true effects sizes in psychological studies based on \cite{van2019}.}
   \label{fig:dist_cohen}
\end{figure}
Figure \ref{fig:dist_cohen} shows the kernel density estimate of the 164 effect sizes (solid line) and a corresponding truncated normal distribution (dashed line). Both distributions would be reasonable choices for prior distributions when calculating $p$ in the thick $t$-test. If the true effects are distributed like this, using a uniform prior to calculate $p$ in the thick $t$-test as we did in the simulation would result in a lower rejection probability of the thick null hypothesis and thus also in a lower false positive rate than $\alpha$ (see Appendix \ref{app:prior_simu} for a related simulation study). 

\subsection{Effect of different prior choices for $t$-test \label{app:prior_simu}}
To test the effect of different prior choices for the thick $t$-test, we ran a second simulation with $\mu$ drawn from $\mathcal{N}(\mu_0, 50 / \sqrt{12})$, where $\mu_0=100$ as before and $\sigma^2 = 50^2 / 12$ is the variance of a continuous uniform distribution on $[75,125]$, the range which $\mu$ is sampled from in the original simulation. As expected, a thick $t$-test with a matching prior distribution, i.e., $\mathcal{N}(\mu_0, 50 / \sqrt{12})$ truncated at the edges of $I_0$, yields a false positive rate of approximately 5\% (Table \ref{tab:error_rates_2}). By contrast, a thick $t$-test with a flat prior, i.e., a continuous uniform distribution on $I_0$, concentrates more probability mass at the edges of $I_0$ and hence assumes $H_0^t$ to be compatible with bigger measured effects than the normal prior. Therefore, $H_0^t$ is rejected less often and the false positive rate is reduced to 4\%. All other methods' false positive rates and other performance measures are changed in a similar way if the relation between their implicit prior assumptions (see Section \ref{sec:methods_t}) and the distribution of $\mu$ in simulation changes. For example, the conventional methods obtain a smaller false positive rate closer to $\alpha$ because the tests' implicitly assumed distribution that places all probability mass on $\mu_0$ is more similar to the normal distribution of the new simulation than the uniform distribution of the original simulation.

\begin{table}[t!]
\caption{Error rates and inference success rate for each method in the simulation setting with $\mu \sim \mathcal{N}(\mu_0, 50 / \sqrt{12})$.}
\label{tab:error_rates_2}
\resizebox{\columnwidth}{!}{
\begin{tabular}{lccccccc}
\toprule
& Conventional & Small-alpha & MESP & Distance-only & Interval-based & Thick $t$-test flat & Thick $t$-test normal\\ 
  \cmidrule(lr){2-8}
  False positive rate & 36.2\% & 22.0\% & 8.8\% & 16.2\% & 0.4\% & 3.8\% & 4.8\% \\ 
  False negative rate & 24.0\% & 37.6\% & 26.9\% & 12.2\% & 53.9\% & 35.7\% & 33.5\% \\ 
  False discovery rate & 34.5\% & 28.1\% & 11.8\% & 17.0\% & 1.0\% & 6.1\% & 7.4\% \\ 
  False omission rate & 25.5\% & 30.5\% & 21.1\% & 11.7\% & 32.9\% & 25.2\% & 24.2\% \\
  \addlinespace
  Inference success rate & 69.6\% & 70.5\% & 82.6\% & 85.6\% & 74.1\% & 81.0\% & 81.5\% \\
  \bottomrule
\end{tabular}
}
\end{table}

\subsection{Effect of different choices of $\alpha$ \label{app:alpha}}
\begin{figure}[t!]
    \centering
    \includegraphics[width=0.48\textwidth]{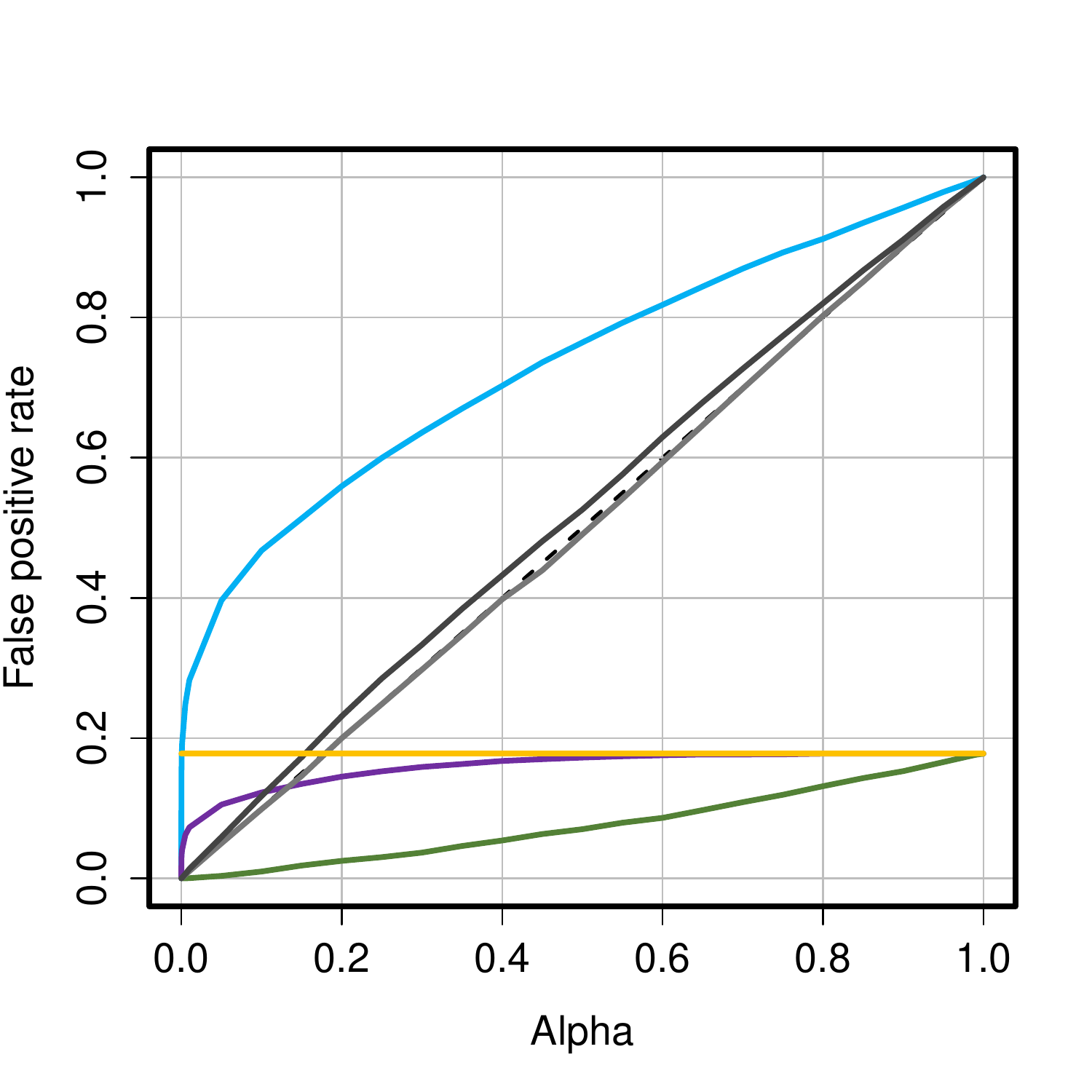}
    \includegraphics[width=0.48\textwidth]{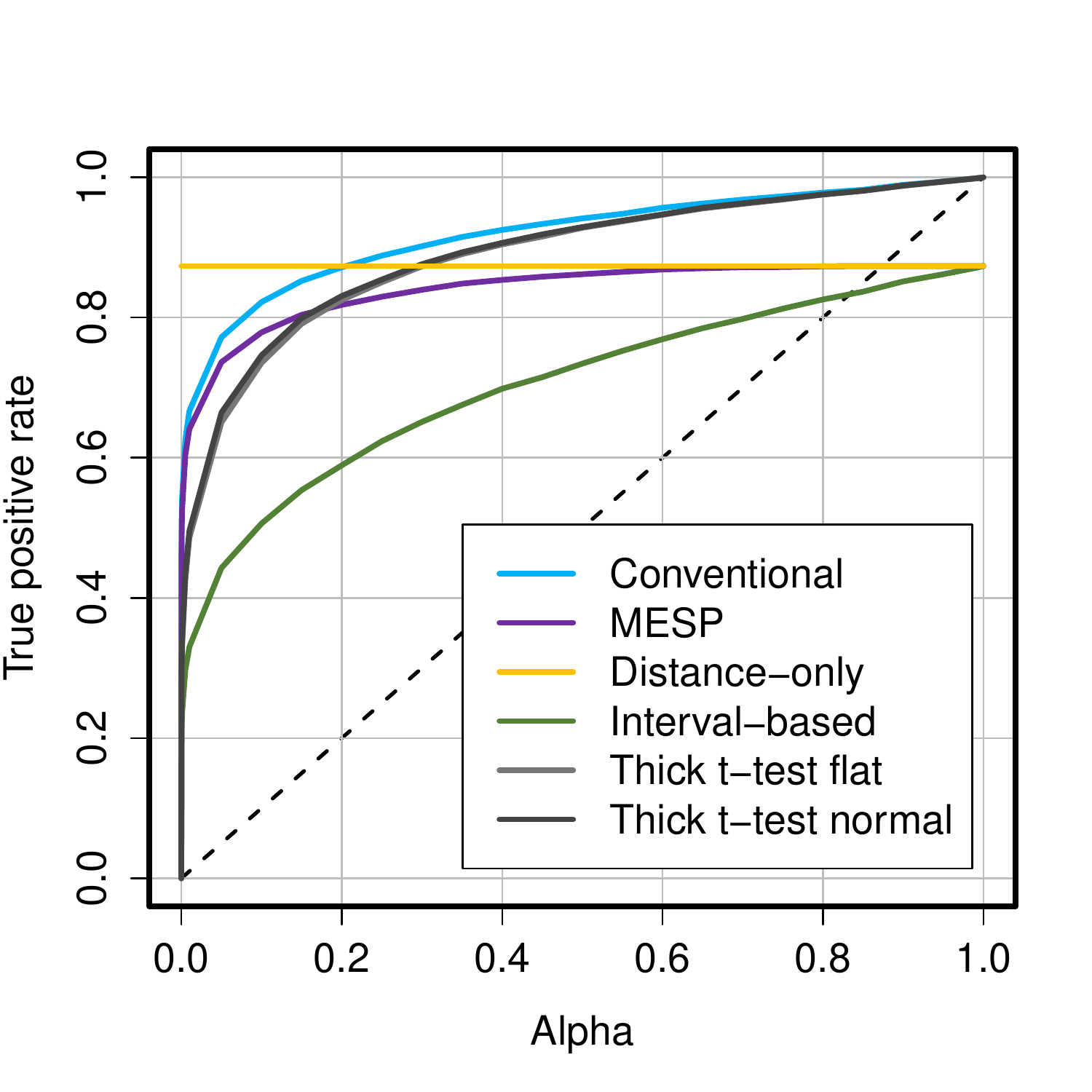}
    \caption{Impact of $\alpha$ on the false and true positive rates.}
    \label{fig:alpha}
\end{figure}

To test the effect of different choices of $\alpha$, we ran a third simulation with $\mu$ drawn from a uniform distribution on $[75,125]$. In every simulated case we tested different $\alpha$ parameters in a range from 0 to 1 for all methods. Figure \ref{fig:alpha} shows that for the thick $t$-test with the prior that matches the simulation distribution, the false positive rate approximately equals $\alpha$ for all choices of $\alpha$. The truncated normal prior concentrates more probability mass on $\mu_0$ and thus, the graph for the thick $t$-test using that prior is shifted slightly towards the direction of the conventional $t$-test with its implicitly assumed prior distribution that places all probability mass on $\mu_0$. We also see that the ability of the MESP and the interval-based method to control their false positive rates and their power using $\alpha$ is limited because they are bound by the distance-only method which provides no $\alpha$ parameter.

\end{document}